\documentclass[onecolumn,showkeys,preprintnumbers,amsmath,amssymb]{revtex4}
\usepackage{graphicx}
\usepackage{dcolumn}
\usepackage{bm}
\usepackage{epsfig}
\usepackage{rotating}

\begin{document}

\title{Hierarchical structure of the countries based on electricity consumption and economic growth}

\author{Ersin Kantar$^{a, b}$, Alper Aslan$^{c}$, Bayram Deviren$^{d}$ and Mustafa Keskin$^{*, a}$
\\
$^{a}$Department of Physics, Erciyes University, 38039 Kayseri, Turkey
\\
$^{b}$Institute of Science, Erciyes University, 38039 Kayseri, Turkey
\\
$^{c}$Faculty of Economics and Business, Nevsehir University,  50300, Nev\c{s}ehir,Turkey
\\
$^{d}$Department of Physics, Nevsehir University, 50300 Nev\c{s}ehir, Turkey
}
\date{\today}

\altaffiliation[]{Corresponding Author: Department of Physics, Erciyes University, 38039 Kayseri, Turkey
\\Email: keskin@erciyes.edu.tr (M. Keskin);
\\Tel.: +90 352 4374938x33105;
\\Fax: +90352 4374931.}

\begin{abstract}
We investigate the hierarchical structures of countries based on electricity consumption and economic growth by using the real amounts of their consumption over a certain time period. We use of electricity consumption data to detect the topological properties of 60 countries from 1971 to 2008. These countries are divided into three subgroups: low income group, middle income group and high income group countries. Firstly, a relationship between electricity consumption and economic growth is investigated by using the concept of hierarchical structure methods (minimal spanning tree (MST) and hierarchical tree (HT)). Secondly, we perform bootstrap techniques to investigate a value of the statistical reliability to the links of the MST. Finally, we use a clustering linkage procedure in order to observe the cluster structure more clearly. The results of the structural topologies of these trees are as follows: i) we identified different clusters of countries according to their geographical location and economic growth, ii) we found a strong relation between energy consumption and economic growth for all the income groups considered in this study and iii) the results are in good agreement with the causal relationship between electricity consumption and economic growth.

\end{abstract}

\keywords{Hierarchical Structure Methods; Bootstrap Technique; Electricity Consumption and Economic Growth}
\maketitle
{JEL classification:} C10, C52, Q40, Q43
\section{\label{sec:level1}Introduction}

Electricity consumption has become a topic of immense importance. The growing interest in developed and developing countries has largely been triggered by the growing demand for energy across the world fueled mainly by increasing economic activities, particularly in emerging countries. Estimating electricity consumption in advance is crucial in the planning, analysis and operation of power systems in order to ensure an uninterrupted, reliable, secure and economic supply of electricity. Moreover, modeling and predicting electricity consumption play a vital role in developed and developing countries for policy makers and
related organizations.

The causal relationship between electricity consumption and economic growth has been investigated and the empirical literature has focused on four hypotheses when dealing with the causal relationship between electricity consumption and economic growth:  conservation, growth, feedback, and neutrality. The first is the conservation hypothesis which is supported if an increase in economic growth causes an increase in electricity consumption. Under this scenario, an increase in economic growth would have a negative impact on electricity consumption.  The second is the growth hypothesis which supposes that electricity consumption can directly impact on economic growth and indirectly as a complement to labor and capital in the production process. The growth hypothesis verified if there is a unidirectional causality from electricity consumption to economic growth. If this is the case, an increase in electricity consumption has a positive impact on economic growth; energy conservation oriented strategies that decrease electricity consumption may have a harmful impact of economic growth. The feedback hypothesis highlights the interdependent relationship between electricity consumption and economic growth. The existence of bidirectional causality between electricity consumption and economic growth provides support for the feedback hypothesis. Fourth, the neutrality hypothesis suggests that energy consumption provides a relatively trivial position in the determination of economic growth.

Payne \cite{Payne2010} compares the various hypotheses associated with the causal relationship between electricity consumption and economic growth using a survey of the empirical literature. The results illustrate that 31.15 \% supported the neutrality hypothesis; 27.87 \% of studies the conservation hypothesis; 22.95 \% the growth hypothesis; and 18.03 \% the feedback hypothesis.

There are several studies in the empirical literature on the causal relationship between electricity consumption and economic growth. The results in the literature, however, are ambiguous and are presented in Table 1.

\begin{table}
\begin{tabular}{ | l | l | p{2.4in} |}\hline
\textbf{Author(s)} & \textbf{Period/Countries } & \textbf{Methodology} \\ \hline
\multicolumn{3}{|c|}{\textbf{Unidirectional causality }} \\ \hline
\multicolumn{3}{|c|}{\textbf{From economic growth to electricity consumption}} \\ \hline
Ghosh \cite{Ghosh2002} & 1950-1997/India  & Johansen-Juselius;Granger causality-VAR \\ \hline
Narayan et al. \cite{Narayan2005} & 1966-1999/Australia & ARDL bounds testing;Granger causality \\ \hline
Yoo and Kim \cite{Yoo2006} & 1971-2002/Indonesia & Engle-Granger;Johansen-Juselius;Hsiao' s causality \\ \hline
Ho and Sui \cite{Ho2007} & 1966-2002/Hong Kong & Johansen-Juselius;Granger causality \\ \hline
Mozumder and Marathe \cite{Mozumder2007} & 1971-1999/Bangladesh  & Johansen-Juselius;Granger causality \\ \hline
Jamil and Ahmad \cite{Jamil2010} & 1960-2008/Pakistan   & Johansen-Juselius;Granger causality \\ \hline
Shahbaz and Feridun \cite{Shahbaz2012}  & 1971-2008/Pakistan & Toda Yamamoto Wald-test causality tests \\ \hline
\multicolumn{3}{|c|}{\textbf{From electricity consumption to economic growth}} \\ \hline
Aqeel and Butt \cite{Aqeel2001} & 1955-1996/Pakistan & Engle-Granger;Hsiao's causality \\ \hline
Shiu and Lam \cite{Shiu2004} & 1971-2000/China & Johansen-Juselius;Granger causality  \\ \hline
Altinay and Karagol \cite{Altinay2005} & 1950-2000/Turkey & Dolado-Lutkepohl test for causality \\ \hline
Lee and Chang \cite{Lee2005} & 1954-2003/Taiwan & Johansen-Juselius;Weak exogeneity test \\ \hline
Yoo \cite{Yoo2005} & 1970-2002/Korea & Johansen-Juselius;Granger causality  \\ \hline
Narayan and Singh \cite{Narayan2007} & 1971-2002/Fiji Islands & ARDL bounds testing;Granger causality \\ \hline
Yuan et al. \cite{Yuan2007} & 1978-2004/China & Johansen-Juselius;Granger causality  \\ \hline
Odhiambo \cite{Odhiambo2009a} & 1971-2006/ Tanzania& ARDL bounds testing;Granger causality-VECM \\ \hline
Abosedra et al. \cite{Abosedra2009} & 1995-2005/ Lebanon & Granger causality \\ \hline
Chandran et al. \cite{Chandran2010} & 1971-2003/Malaysia & ARDL bounds testing;Engle-Granger;Johansen-Juselius;Granger causality \\ \hline
Narayan and Narayan \cite{Narayan2010} & 1980-2006/93 Countries & Granger causality \\ \hline
Ahamad and Nazrul \cite{Ahamad2011} & 1971-2008/Bangladesh & Granger causality \\ \hline
Bildirici and Kayikci \cite{Bildirici2012} & 1990-2009/11 Commonwealth Independent States & Fully Modified Ordinary Least Squares and Panel ARDL \\ \hline
\multicolumn{3}{|c|}{\textbf{Bidirectional causality}} \\ \hline
Yang \cite{Yang2000} & 1954-1997/Taiwan & Engle-Granger;Granger causality-VAR \\ \hline
Jumbe \cite{Jumbe2004} & 1970-1999 Malawi & Engle-Granger;Granger causality-VECM \\ \hline
Zachariadis and Pashourtidou \cite{Zachariadis2007} & 1960-2004/Cyprus &  Johansen-Juselius;Granger causality-VECM \\ \hline
Tang \cite{Tang2008} & 1972-2003/Malaysia & Granger causality \\ \hline
Tang \cite{Tang2009} & 1970-2005/Malaysia & ARDL bounds testing;Granger causality \\ \hline
Odhiambo \cite{Odhiambo2009b} & 1971-2006/South Africa & Johansen-Juselius;Granger \\ \hline
Lean and Smyth \cite{Lean2010} & 1971-2006/Malaysia & A RDL bounds testing;Johansen-Juselius \\ \hline
Ouedraogo \cite{Oue2010} & 1968-2003/Burkina Faso & ARDL bounds testing;Granger \\ \hline
Shahbaz et al. \cite{Shahbaz2011}  & 1971-2009/Portugal & Granger causality \\ \hline
Kouakou \cite{Kouakou2011}  & 1971-2008/Cote d'Ivoire & Granger causality \\ \hline
Gurgul and Lach \cite{Gurgul2011}  & Q1 2000-Q4 2009/Poland  & Toda-Yamamoto \\ \hline
Shahbaz and Lean \cite{ShahbazLean2012}  & 1972-2009/Pakistan & Granger causality \\ \hline
\multicolumn{3}{|c|}{\textbf{Neutrality}} \\ \hline
Wolde \cite{Wolde2006}  & 1971-2001/17 African countries & ARDL Bounds testing;Toda-Yamamoto's causality \\ \hline
Chen et al. \cite{Chen2007} & 1971-2001/10 Asian countries & Johansen-Juselius;Granger causality \\ \hline
Narayan and Prasad \cite{Narayan2008} & 1960-2002/30 OECD countries & Toda-Yamamoto's causality test \\ \hline
Payne \cite{Payne2009} & 1949-2006 US & Granger causality \\ \hline
Ozturk and Acaravci \cite{Ozturk2010} & 1980-2006/4 European countries & ARDL Bounds test and Granger causality \\ \hline
Ozturk and Acaravci \cite{Ozturk2011} & 1971-2006 MENA countries & ARDL Bounds test-VECM \\ \hline
\end{tabular}
\caption{Summary of literature on electricity consumption - economic growth nexus.}
\end{table}

The topic of the causal relationship between energy consumption and economic growth has been well studied in the energy economics
literature. Different studies have focused on different countries, time periods, proxy variables and different econometric methodologies have been used to determine the energy consumption and growth relationship. Moreover, a literature survey on the relationship between energy consumption and economic growth is given in detailed by Ozturk  \cite{Ozturk}.

Complex networks provide a very general framework, based on the
concepts of statistical physics, for studying systems with large
numbers of interacting assets. These networks have been able to
successfully describe the topological properties and characteristics
of many real-life systems such as multilocus sequence typing for
analyses of clonality \cite{Chen2006}, scientific collaboration in
the European framework programs \cite{Garas2008}, taxonomy of correlations
of wind velocity \cite{Bivona2008}, Brazilian term structure
interest rates \cite{Tabak2009}, the international
hotel industry in Spain \cite{Brida2010a}, and foreign trade \cite{Kantar2011}. Moreover, the most recent literature has studied networks generated by correlations of stock prices
\cite{Mantegna1999,Mantegna2000,Bonanno2003,Bonanno2000,Onnela2003a,Onnela2003b,Jung2006,Tumminello2007a,Jung2008,Feng2010,Keskin2010a,Keskin2010b, Kantar2011a,Kantar2011b}. In this paper, we focus on the electricity consumption and the main objective is to
characterize the topology and taxonomy of the network of the countries. To the best of the authors's knowledge, this is the first study on electricity consumption and economic growth by using the hierarchical structure methods.

The aim of the present paper is to examine relationships
among countries, based on low income group, middle income group and high income group countries, by using the concept
of the minimal spanning tree (MST) and hierarchical tree (HT) over the
period between 1971-2008. From these trees, both geometrical (through the
MST) and taxonomic (through the HT) information about the
correlation between the elements of the set can be obtained. Note
that the MST and then the HT are constructed using the Pearson
correlation coefficient as a measure of the distance between the
time series. Moreover, we use the bootstrap technique to associate
a value of reliability to the links of the MST. We also use average linkage cluster analysis to obtain the HT. These
methods give a useful guide to determining the underlying economic
or regional causal connections for individual countries.

The remainder of the paper is structured as follows. The next section briefly
introduces the set of empirical data we work with. Sec. III is targeted at presenting the method. Sec. IV presents the empirical results. Finally, Sec. IV provides some final considerations.

\section{The data}

We chose data on the electricity consumption of 60 low income group, middle income group and high income group countries. We used the data period from 1971 to 2008 and listed the countries and their corresponding symbols in Table 2. The annual amounts were downloaded from the World Bank database (http://data.worldbank.org/).

\section{The method}

In this section, we describe the methodology used for the analysis of the data. Recent empirical and theoretical
analysis have shown that useful economic information can be detected in a correlation matrix using a variety of
methods \cite{Mantegna1999,Mantegna2000,Mizuno2006,Ortega2006,Brida2009a,Feng2010,Keskin2010a,Naylor2007,Bonanno2004,Vandewalle2001,Brida2010,Eom2007,Brida2007,Brida2009b,Garas2007,Brida2010c,Bonanno2000,Coelho2007a,Gilmore2008,Sieczka2009,Tabak2010,Onnela2003a,Onnela2003b,Onnela2002,Onnela2003c,Micciche2003,Coelho2007b}.
In this paper, we use three different approaches, based on hierarchical methods (MST and HT), the
bootstrap technique, and the ALCA technique. We will briefly describe the basic aspects of these three different methods in the subsections.

\subsection{Minimal spanning tree (MST) and hierarchical tree (HT)}

In order to construct the MST following the method suggested by Mantegna \cite{Mantegna1999}, the correlation coefficient between a pair of countries based on electricity consumption should be calculated in the first step. The correlation coefficient between a pair of countries based on electricity consumption defines a degree of similarity between the synchronous time evolution of a pair of assets between the countries.

\noindent
\begin{equation} \label{GrindEQ__1_}
C_{ij} =\frac{\left\langle R_{i} R_{j} \right\rangle -\left\langle R_{i} \right\rangle \left\langle R_{j} \right\rangle }{\sqrt{\left(\left\langle R_{i}^{2} \right. \rangle -\left\langle R_{i} \right\rangle ^{2} \right)\left(\left\langle R_{j}^{2} \right. \rangle -\left\langle R_{j} \right\rangle ^{2} \right)} } ,
\end{equation}

\noindent where ${ R}_{{ i}}$ is the vector of the time series of log-returns, ${ R}_{{ i}} {(t)\; =\; ln\; P}_{{ i}} { (t\; +\; }\tau { )\; -\; ln\; P}_{{ i}} { (t)}$ is the log return, and ${ P}_{{ i}}(t)$ is the electricity consumption amount of a country i (i=1,..., N) at time t. We take $\tau$ as one annual in the following analysis throughout this paper.

We create a country network with a significant relationship between countries using the MST. The MST, a theoretical concept in graph theory \cite{West1996}, is the spanning tree of the shortest length using the Kruskal algorithm \cite{Kruskal1956,Cormen1990,Prim1957}. Hence, it is a graph without a cycle connecting all nodes with links. This method is also known as the single linkage method of cluster analysis in multivariate statistics \cite{Everitt1974}. The MST is generated from the graph by selecting the most important correlations between foreign trade prices. The MST reduces the information space from $\textit{N}(\textit{N - 1})\textit{/2}$ separate correlation coefficients to ($\textit{N - 1}$) linkages, known as tree ``edges'', while retaining the salient features of the system \cite{Gilmore2008}. Therefore, the MST is a tree which has $\textit{N - 1}$ edges that minimize the sum of the edge distances in a connected weighted graph of the \textit{N} rates.

Mantegna \cite{Mantegna1999}, and Mantegna and Stanley \cite{Mantegna2000} showed that the correlation coefficients can
be transformed into distance measures, which can in turn be used to describe hierarchical organization of the group of analyzed assets. Distance measure
\noindent
\begin{equation} \label{GrindEQ__2_}
{\rm d}_{{\rm ij}} =\sqrt{2(1-C_{ij} )} ,
\end{equation}

\noindent where ${\rm d}_{{\rm ij}} $ is a distance for a pair of the rate i and the rate \textit{j}, and it fulfills the three axioms of Euclidean distance \cite{Mantegna1999}.

\noindent Now, one can construct an MST for a pair of countries using the N $\times$ N matrix of ${\rm d}_{{\rm ij}} $. Hence, a country network with a significant relationship between countries using the MST is obtained. The MST, a theoretical concept in graph theory \cite{West1996}, is the spanning tree of the shortest length using the Kruskal algorithm \cite{Kruskal1956,Cormen1990,Prim1957}. Hence, it is a graph without a cycle connecting all nodes with links. This method is also known as the single linkage method of cluster analysis in multivariate statistics \cite{Everitt1974}.

We also introduce the ultrametric distance or the maximal ${\rm d}_{{\rm ij}}^{{\rm \wedge }} $${}_{ }$ between two successive countries encountered in order to construct an HT, when moving from the first country \textit{i} to the last country \textit{j} over the shortest part of the MST connecting the two countries. (For a fuller technical discussion see \cite{Mantegna1999,Mantegna2000,Bonanno2003,Onnela2003a,Tumminello2007a,Feng2010,Keskin2010a,Kantar2011a}.) The hierarchical tree ranks the linkages between countries via the subdominant ultrametric distance, beginning with the pair exhibiting the shortest distance measure. Successive countries are added to the center of this tree in order of increasing distances. Thus, the last country added to the hierarchical tree are those with the most distant linkages to the center country or countries.

\subsection{The stability of links with the bootstrap technique}

The major weakness of the described methodology lies in the fact that the calculated MST and HT might be unstable. Moreover, without further statistical analysis, we cannot be sure whether the links present in the MST are actually the important links in the network or are rather a statistical anomaly, i.e. whether the results are sensitive to the sampling. We use a bootstrap technique proposed by Tumminello et al. \cite{Tumminello2007b,Tumminello2007a,Tumminello2010} specifically for MST and HT analysis to deal with the problem. The bootstrap technique, which was invented by Efron \cite{Efron1979}, and has been widely used in phylogenetic analysis since the paper by Felsenstein \cite{Felsenstein1985} as a phylogenetic hierarchical tree evaluation method \cite{Efron1996}. This technique was used to quantify the statistical reliability of the hierarchical structures of Turkey's foreign trade \cite{Kantar2011} and major international and Turkish companies \cite{Kantar2011a}.

In the technique, by using the original MST and HT, we construct a bootstrapped time series from the original while keeping the the length of the time series fixed (i.e. the observations may repeat in the bootstrapped sample). MST and HT are then constructed for the bootstrapped time series and links are recorded. It is then checked whether the connections in the original MST are also present in the new MST based on bootstrapped time series. We repeat such procedure 1000 times so that we can distinguish whether the connections in the original MST and HT are the strong ones or statistical anomalies \cite{Keskin2010a}. The bootstrap value gives information about the reliability of each link of a graph.

\subsection{Cluster analysis}

The correlation matrix of the time series of a multivariate complex
system can be used to extract information about aspects of the
hierarchical organization of such a system. Correlation based
clustering has been used to infer the hierarchical structure of a
portfolio of stocks from its correlation coefficient matrix
\cite{Mantegna1999,Bonanno2001,Bonanno2003}. The correlation based
clustering procedure also allows a correlation based
network to associate with the correlation matrix. For example, it is natural to
select the MST as the correlation based network associated with
single linkage cluster analysis. A different correlation based on
networks can be associated with the same hierarchical tree putting
emphasis on different aspects of the sample correlation matrix.
Useful examples of correlation based networks apart from the
minimum spanning tree are the planar maximally filtered graph
\cite{Tumminello2005} and the average linkage minimum spanning tree
\cite{Tumminello2007a,Kantar2011,Kantar2011a}.

We use average linkage cluster analysis (ALCA) in order to observe more clearly the different clusters of countries according to their geographical location and economic growth. Since the ALCA, which is a hierarchical clustering method, and an account of the method was presented in detailed by Tumminello et al. \cite{Tumminello2010} and also \cite{Tumminello2007a,Kantar2011,Kantar2011a}, we only give the obtained results. The constructions of the MST and HT will be elaborated in Section IV.

\section{Numerical Results and Discussions}

In this section, we present the MST, including the
bootstrap values, and HT of 60 countries based on electricity consumption
from 1971 to 2008. These countries are divided into three subgroups: low income group, middle income group and high income group countries. We also investigate cluster structures by using a clustering linkage procedure.

We construct the MST by using Kruskal's algorithm
\cite{Kruskal1956,Cormen1990,West1996} for the electricity consumption based on a distance-metric matrix. The amounts of the links that persist from one node (country) to the other correspond to the relationship between the countries in electricity consumption. We carried out the bootstrap technique to associate a value of the statistical reliability to the links of the MST. If the values are close to one, the
statistical reliability or the strength of the link is very high.
Otherwise, the statistical reliability or the strength of the link
is lower \cite{Tumminello2007a,Keskin2010a}. We also obtained the
cluster structure of the hierarchical trees much better by using
average linkage cluster analysis.

Fig. 1 shows the MST applying the method of Mantegna
\cite{Mantegna1999}, Mantegna and Stanley \cite{Mantegna2000} for
electricity consumption based on a distance-metric matrix for the period 1971-2008. In Fig. 1, we observe different clusters of countries according to their geographical proximity and economic
growth. In this figure, we detected three different clusters: mainly European Union countries formed the first cluster of countries with a GDP of over \$30,000; the second cluster was formed mainly by some European and South American countries; and mainly African countries formed the third cluster with a GDP of under \$5,000. It can also be clearly seen that in the MST, the European Union countries form the central structure. It is observed that DEU is at the center of the European Union countries and it is the predominant country for this period. The first cluster consists of DEU, AUT, FRA, ITA, DNK, NLD, ESP, LUX, BEL, IRL, FIN, GBR, SWE, NOR, GRC, USA, JPN, CAN and CHE, which are an European Union countries except USA, JPN, CAN and CHE; hence it is a heterogonous cluster. In this cluster, there are strong relationships among BEL - NLD, SWE - NOR, AUT - CHL and USA - JPN. We can establish this fact from the bootstrap values of the links between these countries, which are equal to 1.00, 0.91, 0.91 and 0.91 in a scale from zero to one, respectively; hence these countries are very closely connected with each other. The second cluster is composed of some European and South American countries, namely, HUN, POL, ROM, BGR, CZE, BRA, ARG, URY, MEX, OMN and NZL. In this cluster, there are strong relationships among HUN - MEX, BRA - OMN and POL - ROM. We can establish this from the bootstrap values of the links among the countries, which are equal to 1.00, 0.87 and 0.78 in a scale from zero to one, respectively. The third cluster was formed by mainly of African countries, and was separated four sub-groups. The first sub-group contains SEN, KEN and ETP, and there is strong relationships between SEN and KEN. We can establish this from the bootstrap value of the link between the SEN and KEN, which is equal to 1.00 in a scale from zero to one. The second sub-group consists of
BEN, BGD and PAK, and the bootstrap values of the links between BEN - BGD and BGD - PAK are equal to 0.83 and 0.74, respectively in this sub-group; hence these countries are very closely connected with each other. (MAR, ZMB and CMR) and (YEM and NPL) formed the third and fourth sub-groups, respectively. On the other hand, the bootstrap values of the links between GHA - ZWE, LUX - CHL, TUR - IND, TUR - VNM and KEN - ETH are very low, as seen in Fig. 1. This means that these links could only demonstrate a statistical fluctuation. It is worth mentioning that in comparison with other regions, such as Latin America, the Middle East, Europe, and North America, Africa has one of the lowest per capita consumption rates. Modern energy consumption in Africa is very low and heavily reliant on traditional biomass.

The HT of the subdominant ultrametric space associated with the MST
is shown in Fig. 2. Two countries (lines) link when a horizontal
line is drawn between two vertical lines. The height of the
horizontal line indicates the ultrametric distance at which the two
countries are joined. To begin with, in Fig. 2, we can observe three clusters. The first cluster is composed of countries with a GDP per capita of over \$30,000 and consists of three sub-groups, namely European Union countries (DEU, AUT, FRA, ITA, BEL, NLD and FIN), USA and JPN, and SWE and NOR. The distance between ITA and BEL is the smallest of the sample, indicating the strongest relationship between these two countries. The second cluster is mainly made up of countries from Europe and South America. In this cluster, the distance between ROU and CZE is the smallest of the sample, indicating the strongest relationship between these two countries. The  third cluster is composed of mainly African countries; it also includes of the two sub-groups, namely YEM and NPL, and PAK and BGD.

In the HT, we used average linkage cluster analysis (ALCA) in order to observe the cluster structure more clearly. The HT seen in Figs. 3 is obtained from data based on electricity consumption for the period 1971-2008. When comparing the HT and ALCA, similar cluster structures were observed; however, the number of countries in the ALCA cluster was found to be more than in the HT. For example, seven countries in the cluster  with a GDP per capita of over \$30,000 were seen in the HT, but seventeen countries were seen in ALCA, as can be verified by comparing Fig. 2 with Fig. 3. In addition the groups of African countries are more clearly. Thus, we see that the cluster structures are obtained more efficiently by using ALCA.

Overall results of the study show that even there is a strong relationship between energy consumption and economic growth for some individual countries, and also three different clusters are detected: mainly European Union countries formed the first cluster of countries with a GDP of over \$30,000; the second cluster was formed mainly by some European and South American countries; and mainly African countries formed the third cluster with a GDP of under \$5,000. In other words, there is an evidence indicating that energy consumption leads economic growth in some of the three income groups considered in this study. Therefore, a stronger energy conservation policy should be pursued in all countries. In addition, policymakers should take into consideration the degree of economic growth in each country when energy consumption policy is formulated.

\section{SUMMARY AND CONCLUSION}

There is a growing literature that examines the relationship
between energy consumption and economic growth. The bulk of this
literature focuses on developing, developed and emerging
countries. It is important for policymakers to understand the
relationship between energy consumption and economic growth
in order to design effective energy and environmental policies. A
general conclusion from these studies is that there is no
consensus either on the existence of the relationship or the
direction of causality between energy consumption and economic
growth in the literature.

In this  paper  attempts  were made to re-examine the strong relationship between
energy consumption and economic growth and vice versa in 60 countries by using the concept of the MST,
including the bootstrap values, and the HT for the 1971-2008 period. We also divided these countries into three subgroups: low income group, middle income group and high income group countries. We obtained the clustered structures of the trees and identified different clusters of countries according to their geographical
proximity and economic growth. From the topological structure of these
trees, we found that the European Union countries are at the center of
the network and the bootstrap values show that they are closely connected to each other.
We also found that these countries play an important role in
world electricity consumption. Moreover, African countries have low energy consumption compared to other regions such as Latin America, the Middle East, Europe, and North America. We performed the bootstrap technique to associate a value of statistical reliability to the links of MST to obtain information about
the statistical reliability of each link of the trees. From the results of
the bootstrap technique, we can see that, in general, the bootstrap
values in the MST are highly consistent with each other. We
also used average linkage cluster analysis to obtain the cluster
structure of the hierarchical trees more clearly. The results are in good agrement with the causal relationship between the electricity consumption and economic growth along a survey of the empirical literature. The findings of this study have important policy implications and it shows that this issue still deserves further attention in future research. Finally, it is important for policymakers to understand the
relationship between energy consumption and economic growth in order to design effective energy and environmental policies.

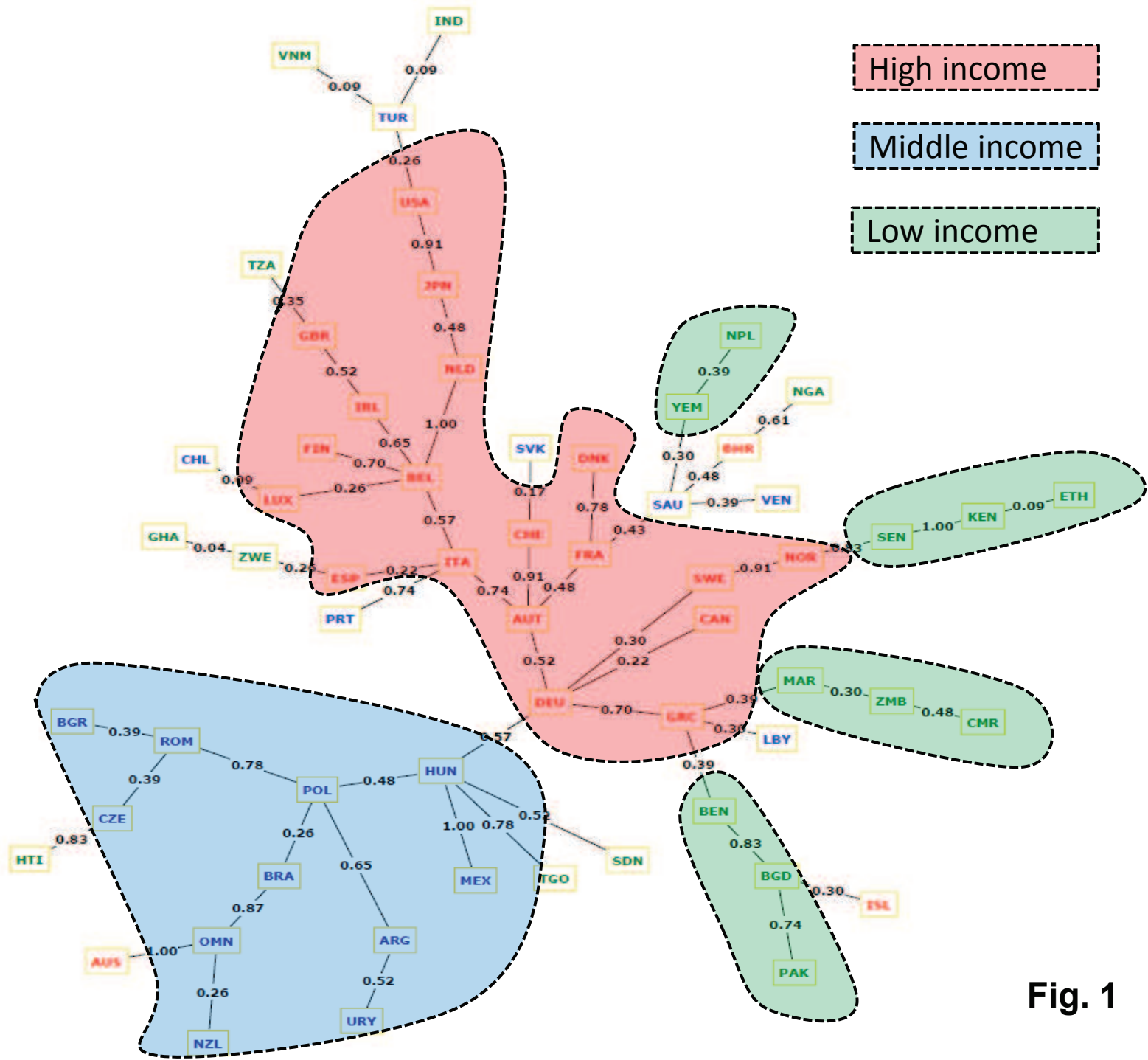
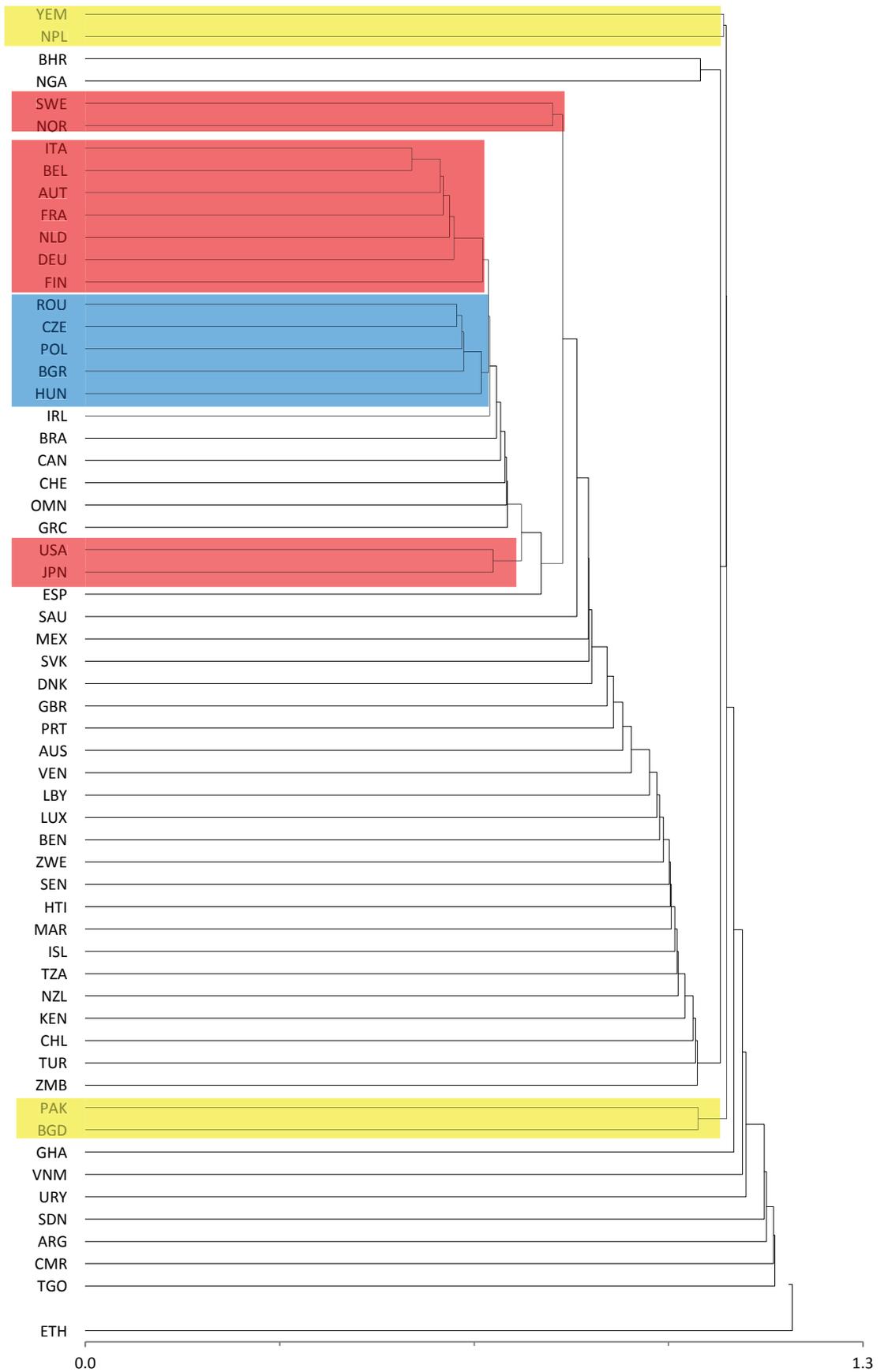
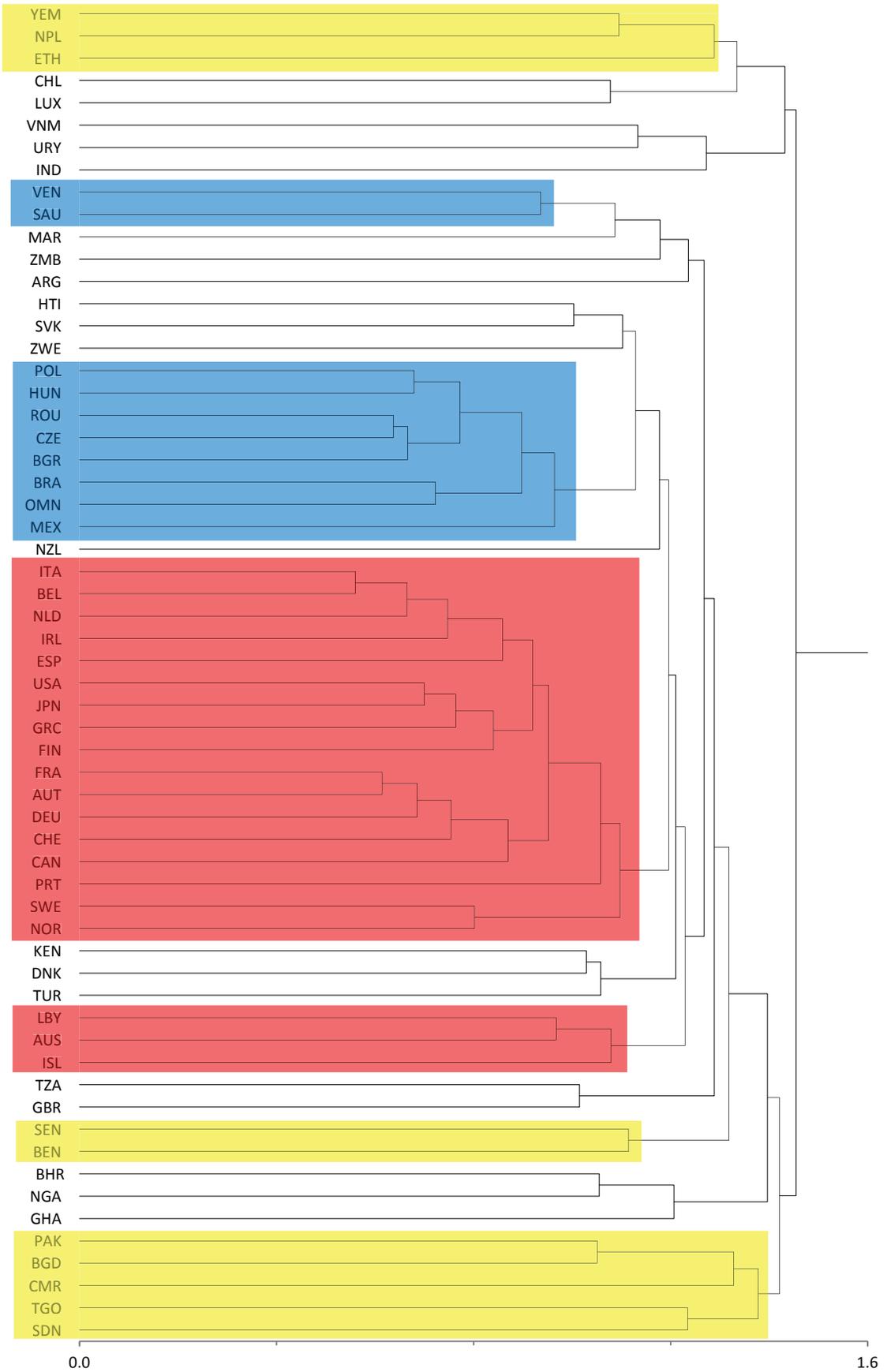
\end{document}